\documentclass[
  aps,
  prb,
  twoside,
  twocolumn,
  showpacs,
  floatfix,
  superscriptaddress,
  10pt
]{revtex4-1}

\usepackage{graphicx}
\usepackage{amsmath}
\usepackage{subfigure}
\usepackage{color}

\newcommand{\etal}[0]{\textit{et al.}}

\newcommand{\eV}[0]{\text{eV}}

\newcommand{\AAA}[0]{\text{\AA}}

\newcommand{\sect}[1]{Sect.~\ref{#1}}
\newcommand{\fig}[1]{Fig.~\ref{#1}}
\newcommand{\eq}[1]{Eq.~(\ref{#1})}

\renewcommand{\vec}[1]{\ensuremath\boldsymbol{#1}}
\renewcommand{\epsilon}[0]{\varepsilon} 
\newcommand{\braket}[3]{\bigl{<}#1\big|#2\big|#3\bigr{>}}

\newcommand{\bra}[1]{\bigl{<}#1\big|}
\newcommand{\ket}[1]{\big|#1\bigr{>}}

\begin{document}


\title{
  Electronic structure of LaBr$_3$ from quasi-particle self-consistent $GW$ calculations
}
\author{Daniel \AA{}berg}
\email{aberg2@llnl.gov}
\affiliation{
  Physical and Life Science Directorate, Lawrence Livermore National Laboratory, California, USA
}

\author{Babak Sadigh}
\affiliation{
  Physical and Life Science Directorate, Lawrence Livermore National Laboratory, California, USA
}

\author{Paul Erhart}
\email{erhart@chalmers.se}
\affiliation{
  Physical and Life Science Directorate, Lawrence Livermore National Laboratory, California, USA
}
\affiliation{  
  Department of Applied Physics, Chalmers University of Technology, Gothenburg, Sweden 
}

\begin{abstract}
Rare-earth based scintillators in general and lanthanum bromide (LaBr$_3$) in particular represent a challenging class of materials due to pronounced spin-orbit coupling and subtle interactions between $d$ and $f$ states that cannot be reproduced by standard density functional theory (DFT). Here a detailed investigation of the electronic band structure of LaBr$_3$ using the quasi-particle self-consistent $GW$ (QPscGW) method is presented. This parameter-free approach is shown to yield an excellent description of the electronic structure of LaBr$_3$. Specifically it is able to reproduce the band gap, the correct level ordering and spacing of the $4f$ and $5d$ states, as well as the spin-orbit splitting of La-derived states. The QPscGW results are subsequently used to benchmark several computationally less demanding techniques including DFT+$U$, hybrid exchange-correlation functionals, and the $G_0W_0$ method. Spin-orbit coupling is included self-consistently at each QPscGW iteration and maximally localized Wannier functions are used to interpolate quasi-particle energies. The QPscGW results provide an excellent starting point for investigating the electronic structure of excited states, charge self-trapping, and activator ions in LaBr$_3$ and related materials.
\end{abstract}

\maketitle

\section{Introduction}

Scintillators are materials that exhibit luminescence upon excitation by ionizing radiation, \cite{Rod97} which means that a fraction of the absorbed energy is re-emitted as light. The emitted photons can be subsequently converted into an electric current using for example photomultiplier tubes or photo diodes, which allows one to measure the energy spectrum of the incoming radiation. Scintillation is observed in crystals, plastics, liquids, and glasses. \cite{Kno10} Examples of inorganic crystal scintillators include halides, oxides, and chalcogenides. \cite{Rod97, Kno10} The emitted light can be produced by exciton recombination (e.g., alkali halides, CsI, MgWO$_4$), core-to-valence, also known as cross-luminescence or Auger-free luminescence, (e.g., BaF$_2$, CsCl) or most commonly by the relaxation of an excited activator atom (e.g., NaI:Tl, SrI$_2$:Eu, LaF$_3$:Ce). \cite{Rod97}

Over the course of the last couple years, interest in scintillator materials has surged thanks to large scale applications in nuclear and radiological surveillance, high-energy physics and medical imaging. While the general potential of scintillators has been demonstrated, one of the current goals is to develop materials with improved energy resolution sufficient to detect fissile materials with a low probability of errors at ports, borders, and airports. \cite{NelGosKna11} The current state-of-the-art material is Ce-doped LaBr$_3$, \cite{LoeDorEij01} which has been extensively characterized both experimentally \cite{DorLoeVin06, DotMcGHar07, Dor10, BizDor07} and theoretically. \cite{BizDor07, VanJafKer10, CanChaBou11, AndKolDor07, Sin10, McIGaoTho07} Yet fundamental features of its electronic structure are still incompletely described and quantified. Such information is, however, crucial for understanding the much improved performance of LaBr$_3$ compared to other scintillators. 

An {\it ab-initio} description of the electronic structure of LaBr$_3$ is challenging due to the presence of La-$4f$ and $5d$ states as well as pronounced spin-orbit coupling (SOC). If one furthermore aims to model Ce activator ions, the Ce-derived $f$ and $d$ levels need to be considered as well. 

Previous electronic structure calculations of LaBr$_3$ have been based on the Hartree-Fock (HF) method, \cite{VanJafKer10}, the LDA+$U$ approach, \cite{CanChaBou11} or hybrid exchange-correlation (XC) functionals. \cite{VanJafKer10} As is well-known, the HF method grossly overestimates the band gap in extended systems and the two latter methods both rely on additional fitting parameters. They are therefore not predictive and require experimental information or higher-level calculations as reference. In view of this situation the objective of the present work is to calculate the electronic structure, specifically the quasi-particle (QP) spectrum, of lanthanum bromide from an essentially parameter-free approach. In doing so one obtains a very good starting point for investigating e.g., charge self-trapping, exciton spectra or activator level alignment. To accomplish this goal the quasi-particle self-consistent $GW$ (QPscGW) method \cite{FalSchKot04, SchKotFal06, KotSchFal07} is employed in conjunction with maximally localized Wannier functions. \cite{MarVan97, SouMarVan01}

The remainder of this paper is organized as follows. In the next section the methodology and computational details are reviewed. Section~\ref{sect:results_qpscgw} presents as the main result of the present work the band structure and density of states from QPscGW calculations including SOC. The convergence of our calculations is elaborated in \sect{sect:results_convergence}. A detailed comparison with both hybrid, DFT+$U$, and $G_0W_0$ calculations is presented in \sect{sect:results_hybrid}. Finally, the main conclusions and an outlook of future work are given in \sect{sect:conclusions}.

\section{Methodology}
\label{sect:methodology}

\subsection{The $GW$ approximation}

In the most common implementation of the $GW$ approximation\cite*{Hed65, HedLun70, AryGun98, AulJonWil00} QP energies are computed to zeroth order in perturbation theory according to
\begin{align}
  \epsilon_{n\vec{k}}
  &= \epsilon_{n\vec{k}}^0 + \mathcal{Z}_{n\vec{k}}
  \label{eq:gw}
  \\
  &\quad
  \text{Re} \left[
    \left< \psi_{n\vec{k}} | T + V_{n-e} + V_H
    + \Sigma(\epsilon_{n\vec{k}}^0)|\psi_{n\vec{k}} \right>
    - \epsilon_{n\vec{k}}^0
    \right].
  \nonumber
\end{align}
Here, $T$, $V_{n-e}$, and $V_H$ denote the kinetic energy term as well as the nucleus-ion and Hartree potentials, respectively. The renormalization factor (or QP weight) $\mathcal{Z}_{n\vec{k}}$ is obtained from the energy derivative of the self-energy. \cite{ShiKre07} Typically, the unperturbed single particle energies and wave functions are obtained from a density functional theory (DFT) or HF calculation. This approach is usually referred to as the $G_0W_0$ method.

Obviously, the application of \eq{eq:gw} does not alter the underlying single particle orbitals. It is primarily for this reason that a starting guess is usually required to yield qualitatively the correct level ordering (at least of the occupied states). This condition is violated for example in lanthanide oxides and can be overcome by generating the initial wave function and QP energies using the LDA+$U$ method. \cite{JiaGomRin09} From this description it is apparent that the $G_0W_0$ method yields results that are dependent on the initial wave function.

To overcome these limitations Kotani, Schilfgaarde, and Faleev developed the QP self-consistent $GW$ method. \cite{FalSchKot04, SchKotFal06, KotSchFal07} The key idea in the QPscGW approach is to optimize an effective self-consistent non-interacting Hamiltonian $H^0=T+V_\text{eff}$ such as to reproduce the energy dependence of the self-energy as closely as possible within the random phase approximation (RPA). This is accomplished by minimizing a norm that measures the difference between $H^0$ and $H(\omega)$ with respect to the effective one-body potential $V_\text{eff}$. A practical scheme is obtained by requiring the XC potential to depend on an average of the {\em Hermitian} part of the self-energy operator,
\begin{align}
  V_{xc} &= \frac{1}{2} \sum_{ij} \left|\psi_i\right>
  \left\{ \text{Re}\left[\Sigma(\epsilon_i)\right]_{ij} 
  +       \text{Re}\left[\Sigma(\epsilon_j)\right]_{ij} \right\}
  \left<\psi_j\right|,
\end{align}
where Brillouin zone indices have been dropped for brevity. It can be argued that within the limits of the RPA the solutions of $H^0$ can be interpreted as quasi-particles. \cite{SchKotFal06}

The original form of the QPscGW method was later slightly modified by Shishkin, Marsman, and Kresse \cite{ShiMarKre07} and extended to account for vertex correction in $W$. For practically all materials considered so far the QPscGW has been found to give band gaps in very good agreement with experiment safe for a slight overestimation in particular for small gap materials. \cite{ChaSchKot06}
\footnote{
  It has been suggested that somewhat larger deviations for example for C are due to effect of electron-phonon coupling on the band gap (Ref.~\onlinecite{GonBouCot11}).
}

From our point of view the two major features of the QPscGW method in comparison to the $G_0W_0$ method described by \eq{eq:gw} are that ({\it i}) the single particle orbitals are updated during the course of the self-consistency loop and ({\it ii}) the final results are independent of the initial wave function and thus also independent of any adjustable parameters. For the latter statement to be fulfilled, it is essential that the basis in which $H^0$ is expanded is sufficiently large to be considered complete as discussed in \sect{sect:results_convergence}.

\subsection{Maximally localized Wannier functions}
Wannier functions \cite{Wan37} (WFs) provide a complementary basis set to the Bloch functions. WFs are defined as Fourier transformations of Bloch functions with respect to crystal momentum vectors. This formulation is, however, arbitrary due to the undetermined phases of the Bloch functions. This was utilized by Marzari and Vanderbilt who introduced generalized Wannier functions\cite{MarVan97} for composite bands defined by
\begin{align}
  w_{\vec{R}i}(\vec{r}) &= \frac{\Omega_\text{cell}}{(2\pi)^3}
  \int_\text{BZ} d\vec{k} e^{-i\vec{k}\cdot\vec{R}}
  \sum_{n=1}^{N_\mathbf{k}} U^{(\vec{k})}_{ni} \psi_{\vec{k}i}(\vec{r}).
\end{align}
Here the unitary matrices $U^{(\vec{k})}$ mix Bloch states having the same wave vector $\vec{k}$. The quadratic spread of the set of WFs can then be expressed in terms of matrix elements
\begin{align}
  M^{(\vec{k},\vec{b})}_{mn}
  = \braket{\psi_{\vec{k}m}}{e^{-i\vec{b}\cdot\vec{r}}}{\psi_{\vec{k}n}},
  \label{eq:mlwf}
\end{align}
and subsequently minimized to obtain so-called maximally localized Wannier functions (MLWFs). This formalism was later extended by Souza {\em et al.} to the case of entangled energy bands. \cite{SouMarVan01} For the purpose of the present work, the underlying first-principles code was modified to enable the calculation of the matrix elements appearing in \eq{eq:mlwf}, \cite{AbeErhCro10} which were then used as input for \textsc{wannier90} \cite{MosYatLee08} to obtain MLWFs. The site and angular momentum projected QP energies were finally interpolated using the MLFW basis to generate accurate band structures and DOS. 

\subsection{Spin-orbit coupling and symmetry operations}

The spin-orbit correction to the Hamiltonian can be obtained from the Dirac equation by three successive Fouldy-Wouthuysen transformations. \cite{BjoDre64} In atomic units, the resulting Hamiltonian for an electron in a central potential $V$ can be written as
\begin{align}
  H_{\text{so}}&= \frac{\alpha^2}{2}\frac{1}{r}\frac{dV}{dr} \vec{\ell}\cdot\vec{s}.
  \label{eq:so}
\end{align}
Here $\alpha$ is the fine structure constant and $\vec{\ell}$ and $\vec{s}$ are the orbital and spin angular momentum operators, respectively. Aryasetiawan and Biermann generalized Hedin's equations\cite*{Hed65, HedLun70} to the case of spin-dependent interactions. \cite{AryBie08,AryBie09} The algebraic structure of the modified set of equations is very similar to Hedin's original equations, although it is shown that in general the self-energy becomes spin-dependent and the polarization now describes the response of the charge density to magnetic fluctuations and vice versa. In the absence of explicit two-particle spin-interaction the self-energy will only depend on variations in the electric field. \cite{AryBie08} Therefore, if vertex corrections, spin-spin, and spin-other-orbit interactions are neglected, this reduces for a non-magnetic system to adding $H_{\text{so}}$ to the unperturbed single-particle Hamiltonian and embodies an explicit spin dependence of the Green's function. Sakuma {\em et al.}\cite{SakFriMiy11} applied this formalism to study the effect of SOC in Hg chalcogenides in the $G_0W_0$ approximation and found an enhancement of the spin-orbit splitting of 0.1\, eV. In the context of QPscGW calculations Chantis {\em et al.} have included SOC as a perturbation on top of already converged orbitals for III-V and II-VI zincblende semiconductors.\cite{ChaSchKot06} In the present case, to investigate spin-orbit splitting SOC is included in the {\em self-consistency} cycles of the QPscGW method. 

In the projector augmented wave method \cite{Blo94, KreJou99} (PAW) the SOC Hamiltonian consists of three terms
\begin{align}\
  \tilde H_{\text{so}} &= H_{\text{so}} + \sum_{ij}  
  \ket{\tilde p_i}
  \left(
    \braket{\phi_i}{H_{\text{so}}}{\phi_j} -
    \braket{\tilde \phi_i}{H_{\text{so}}}{\tilde \phi_j} 
  \right)
  \bra{\tilde p_j},
  \label{eq:paw}
\end{align}
where the projectors $\tilde p$ and orbitals $\phi$ and $\tilde \phi$ have their usual meaning. \cite{Blo94} In DFT calculations it is customary to only include the SOC in a sphere around each atom, using the spherically averaged self-consistent Kohn-Sham potential. Assuming a complete PAW basis set, the first and third terms in \eq{eq:paw} cancel. In the current implementation only the second term is kept.\cite{Mar11} 

For the QPscGW calculations reported in this work the local potential in \eq{eq:so} is computed at each step in the self-consistent cycle as the spherically averaged sum of the external, Hartree and PBE \cite{PerBurErn96} XC potential. \cite{Mar11} As a result, these calculations include SOC only on a DFT level.

To render hybrid XC and $GW$ calculations including SOC computationally feasible it is imperative to take advantage of the space group symmetries and solve for QPs within the irreducible Brillouin zone only. To this end we augmented the action of a symmetry operator in Cartesian space on wave functions, projected wave functions, and the action of the self-energy operator and its derivative with respect to energy on a Bloch function to include rotations in spin space. Now, a general space group operation contains a point group operator $S$ and a translation $\vec{w}$. The action of this operator on a spinless Bloch function is\cite{BraCra72}
\begin{align}
  \{S|\vec{w}\}\exp &
  \left[ iS\vec{k}\cdot\vec{r}\right] u_{\vec{k}}(\vec{r})  =  \nonumber \\
   & \exp
  \left[ iS\vec{k}\cdot(\vec{r}-\vec{w})\right] u_{S\vec{k}}(\vec{r}-\vec{w}),
  \label{eq:spaceop}
\end{align}
where $u_{\boldsymbol{k}}$ is the cell-periodic wave function. The rotated wave function thus corresponds to a Bloch function of wave vector $S\vec{k}$. If the point group operator $S$ contains a rotation by an angle $\alpha$ around some unit axis $\hat{\vec{n}}$ the action of the space group operator on a Bloch spinor would be augmented by an additional rotation of the two spin components by the spin-space rotation matrix
\begin{align}
  R_s(\hat{\vec{n}}, \alpha)
  &= \exp \left[
    -\frac{i\alpha \boldsymbol\sigma \cdot \hat{\vec{n}} }{2} \right]
  = \exp (A)
\end{align}
where $\sigma_i$ is a Pauli matrix. In practice this rotation matrix is computed by $R_s=V^\dagger \exp(D) V$ where $V$ and $D$ are the eigenvector matrix and diagonal eigenvalue matrix of $A$, respectively.

We also note in passing that operations of the type given by \eq{eq:spaceop} coupled with spin rotations are also useful for the evaluation of the matrix $M^{(\vec{k},\vec{b})}$ defined in \eq{eq:mlwf}.

\subsection{Computational details}

Calculations have been performed using the projector augmented wave method \cite{Blo94, KreJou99} as implemented in the Vienna ab-initio simulation package. \cite{KreHaf93, KreHaf94, KreFur96a, KreFur96b}
First wave functions and QP energies were obtained within a generalized Kohn-Sham scheme using several different approximations to represent exchange and correlation effects. Subsequently the single particle states of these calculations served as starting points for $G_0W_0$ and QPscGW calculations. For the XC potential in the initial calculations, we considered a generalized gradient functional, \cite{PerBurErn96} the DFT+$U$ method \cite{LieAniZaa95} with parameters from Ref.~\onlinecite{CzySaw94}, several range-separated hybrid XC functionals with variable mixing parameter $\alpha$ and a fixed screening length of $\mu=0.2\,\AAA^{-1}$ , \cite{HeyScuErn03} as well as exact-exchange (EXX), where the latter is equivalent to carrying out a restricted HF calculation.

All calculations were performed at the experimental lattice parameters \footnote{
  Lanthanum bromide adopts a hexagonal lattice structure in space group 176 (P$6_3$/m) with lattice constants $a=7.9648(5)\,\AAA$ and $c=4.5119(5)\,\AAA$. Lanthanum ions occupy Wyckoff sites $2c$ while Br ions sit on Wyckoff sites $6h$ with $x=0.38506(6)$ and $y=0.29878(6)$ (Ref.~\onlinecite{KraSchSch89}).
}
using a $\Gamma$-centered $6\times 6\times 3$ grid, a general plane-wave cutoff energy of 219\,eV, and a cutoff energy of 146\,eV for the response function in the $GW$ loop. The Green's function and the screened interaction in the $GW$ loop were evaluated using 2088 and 1056 bands in calculations with and without SOC, respectively. The effective QPscGW Hamiltonian was expanded in a basis containing up to 192 bands (384 when including SOC) equivalent to states up to 34\,eV above the conduction band minimum (CBM). In general, our parameters ensure convergence of the valence band and lower conduction band QP energies of about $0.05\,\eV$. The convergence of our calculations is demonstrated and discussed in \sect{sect:results_convergence}.

\section{Results}
\label{sect:results}

\subsection{Band structure and density of states}
\label{sect:results_qpscgw}

\begin{figure}
  \centering
\includegraphics[scale=0.62]{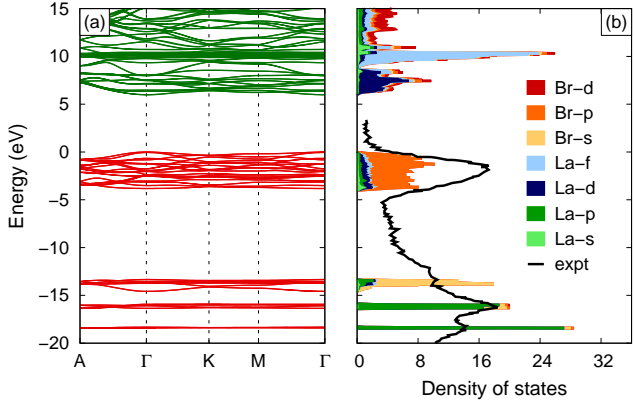}
  \caption{
    (a) Band structure and (b) density of states of LaBr$_3$ as obtained from QPscGW calculations including spin-orbit coupling. Experimental X-ray photoelectron spectrum from Ref.~\onlinecite{Sat76}.
  }
  \label{fig:band_qpscgw}
\end{figure}

Figure~\ref{fig:band_qpscgw} represents the main result of the present paper. It shows the band structure and density of states (DOS) of lanthanum bromide from QPscGW calculations taking SOC into account. As in virtually all halides that we are aware of, the valence band is predominantly composed of halogen $p$-states. Similar to other compounds with comparably low symmetry and a valence band maximum (VBM) that is primarily derived from $p$-states (for example In$_2$O$_3$, Refs.~\onlinecite{ErhKleEgd07, WalSilWei08}, and SrI$_2$, Ref.~\onlinecite{Sin08}), the VBM in LaBr$_3$ is very flat indicating a very low hole mobility.

By comparison the conduction band structure is more complex. Its bottom mostly consists of La-$5d$ states that extend up to about 3.0\,eV above the CBM where a minimum in the DOS is observed. At slightly higher energies a pronounced peak due to La-$4f$ states is clearly visible, superimposed on a rather broad band that has predominantly Br-derived $d$-character.

\begin{figure}
  \centering
\includegraphics[scale=0.62]{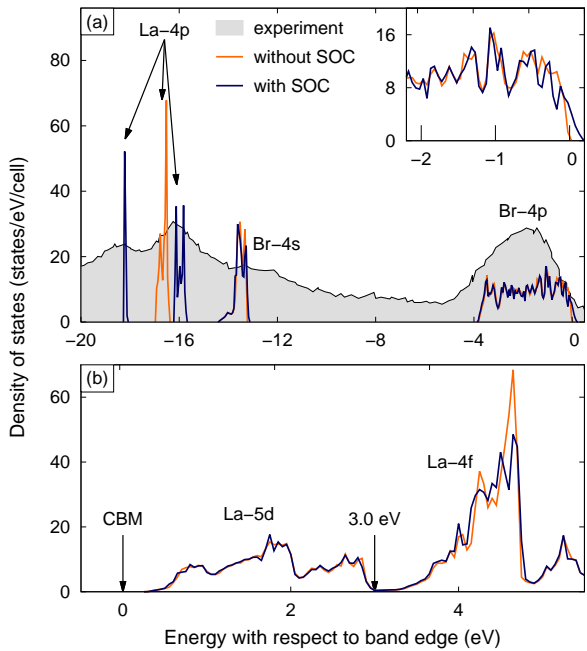}
  \caption{
    Comparison of (a) valence and (b) conduction band density of states from QPscGW calculations with and without spin-orbit coupling (SOC). Valence and conduction band states are shown with respect to valence band maximum and conduction band minimum, respectively obtained from calculations without SOC to highlight the SOC-induced shift.
  }
  \label{fig:soc}
\end{figure}

The SOC lifts the top of the valence band by 0.19\,eV while leaving both the bottom of the Br-$4p$ band and the Br-$4s$ band unchanged. The most significant effect is observed for the La-$4p$ derived states which split by 2.25\,eV in excellent agreement with experiment [compare \fig{fig:soc}(a)]. In general over the energy considered the agreement of the QPscGW calculation with X-ray photoelectron spectroscopy (XPS) data \cite{Sat76} is very good if SOC is included.

The bottom of the conduction band, which is composed of La-$5d$ states, is virtually unaffected by SOC. There is also no change in the position of the La-$4f$ peak situated about 3.2\,eV above the CBM, although one notices a slight redistribution of weights in this band. The QPscGW calculations yield a band gap of 6.19\,eV, which is reduced to 5.99\,eV when SOC is taken into account. This value is in very good agreement with experiments, which indicated a value of 5.9\,eV. \cite{DorLoeVin06}

\begin{figure}
  \centering
\includegraphics[scale=0.62]{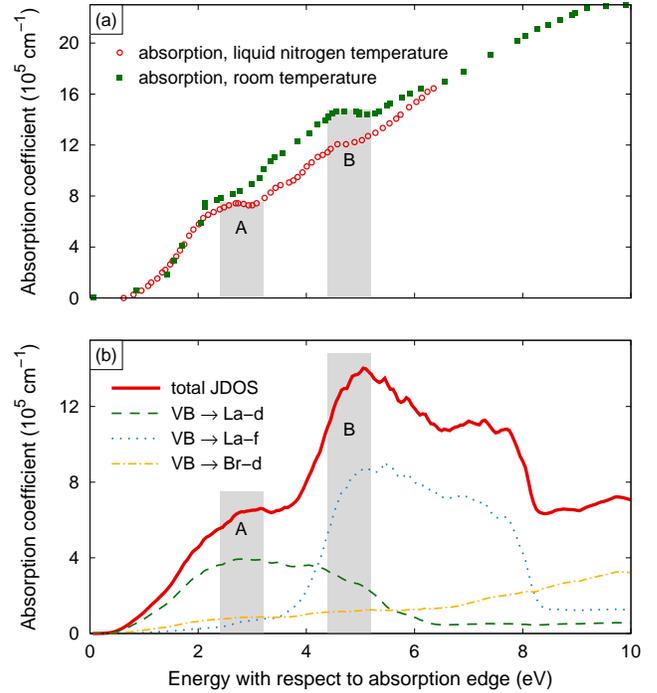}
  \caption{
    (a) Absorption spectra measured by Sato \cite{Sat76} and (b) state resolved contributions to joint density of states (JDOS) from QPscGW calculations due to transitions from valence band (VB) states to different conduction states (SOC contributions are included).
  }
  \label{fig:jdos_abs}
\end{figure}

To the best of our knowledge, there are currently no spectroscopic data corresponding to the single-particle excitation spectrum that would enable a direct comparison of the calculated conduction band structure with experiment. Sato \cite{Sat76} measured the absorption spectrum, which depends on both valence and conduction band states. Since excitonic effects are known to be important in halides, as shown specifically for LaBr$_3$ in Ref.~\onlinecite{DorLoeVin06}, an accurate description of absorption would require the inclusion of electron-hole interactions (see e.g., Refs.~\onlinecite{BenShiBoh98, MaRoh07}). This could be accomplished for example by solving the Bethe-Salpeter equation, \cite{SalBet51, BenShi99, OniReiRub02, MaRoh07, FucRodSch08} which is, however, the subject of future work. In the present paper, we resort to a simplified comparison based on the joint density of states (JDOS). On a single-particle level, the latter is related to the dielectric function and thus the absorption coefficient as discussed for example, in Ref.~\onlinecite{SadErhAbe11}. One can therefore expect a correlation between characteristic features in the absorption spectrum with features in the JDOS.

As observed in \fig{fig:jdos_abs}, there indeed exist similar features in absorption spectra measured by Sato \cite{Sat76} and the JDOS obtained from QPscGW calculations, as indicated by letters A and B.

When separating the contributions from different conduction band states to the JDOS, Figure~\ref{fig:jdos_abs}(b) shows that features A and B arise from transitions from valence band (VB) to La-$d$ and $f$ states, respectively. The fact that the computed features occur at slightly higher energies compared to experiment is consistent with the observation that electron-hole interactions usually lead to a red shift of the calculated spectrum. In \sect{sect:results_hybrid} we compare the JDOS obtained from different computational methods, which will demonstrate that reproducing the two features discussed above is not trivial. The good agreement between the QPscGW JDOS and the XPS data therefore provides evidence for the reliability and accuracy of the QPscGW results.

\subsection{Convergence of QPscGW calculations}
\label{sect:results_convergence}

\begin{figure}
  \centering
\includegraphics[scale=0.62]{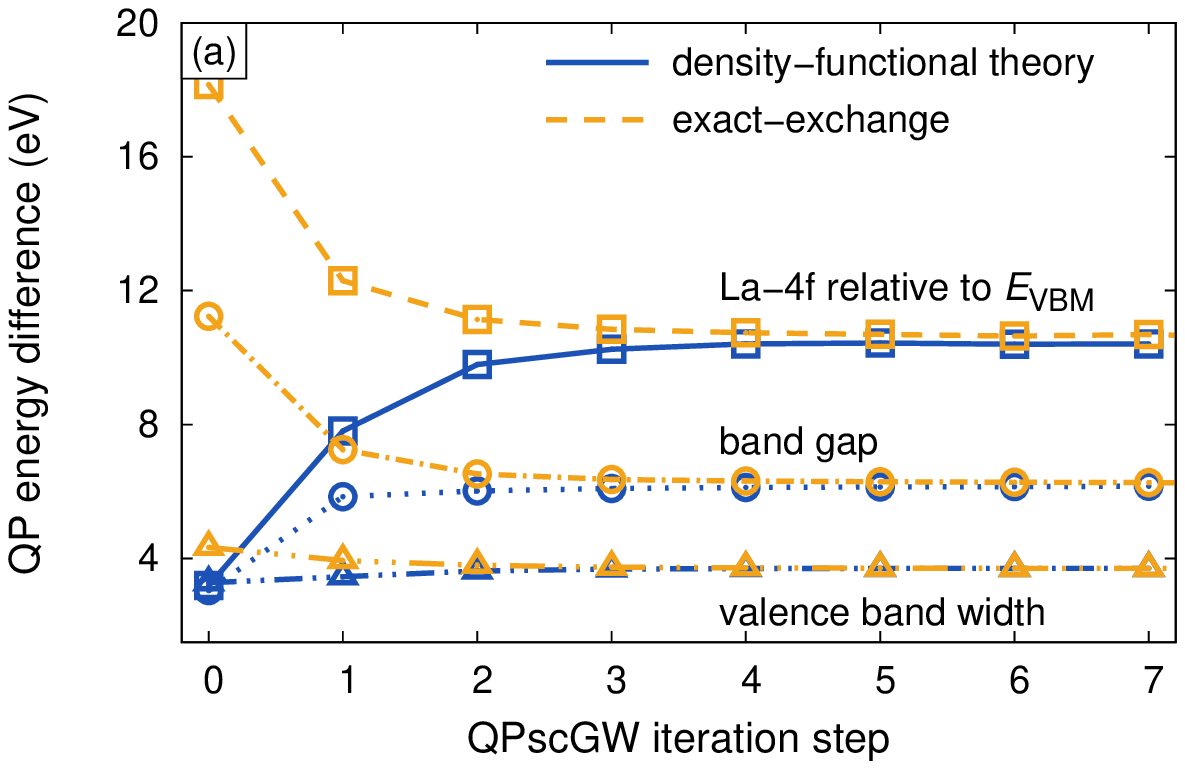}
\includegraphics[scale=0.62]{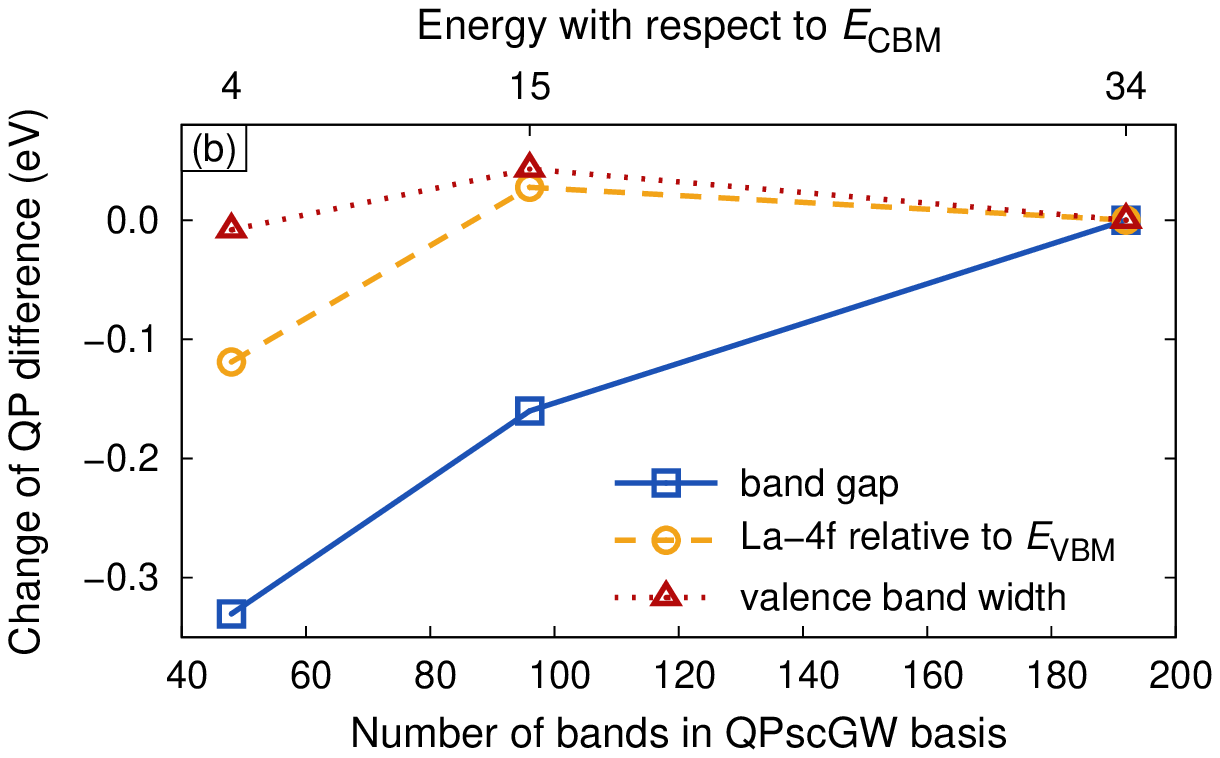}
  \caption{
    (a) Convergence of QPscGW calculations starting from DFT and EXX wave functions, respectively, as evidenced by several quasi-particle energy differences. (b) Convergence with respect to QPscGW basis set size starting from DFT wave function. The results shown in (a) were obtained using 192 bands to expand the effective QPscGW Hamiltonian, corresponding to states up to 34\,eV above the CBM. The calculations shown do not include SOC.
  }
  \label{fig:qp_conv}
\end{figure}

In \sect{sect:methodology} it was argued that the QPscGW method yields results that are independent of the starting wave function. This feature is illustrated in \fig{fig:qp_conv}(a), which compares the convergence of several characteristic QP energy differences starting from both DFT and EXX wave functions. The two different starting points correspond to a significant underestimation (DFT) and overestimation (EXX) of the band gap, respectively. Yet after converging the QPscGW cycles, the converged QP energy differences agree to within 0.05\,eV for valence and lower conduction band levels, and to 0.18\,eV for the La-$4f$ states. The remaining differences can in principle be reduced further by increasing the basis set in which the effective Hamiltonian of the QPscGW method is expanded.

The previous statement is confounded by \fig{fig:qp_conv}(b), which shows the convergence of several QP energy differences with respect to the number of bands included in the QPscGW basis set. \footnote{
  Note that the number of bands included in the QPscGW basis is distinct from the number of bands used to construct $G$. For the present system the latter is usually more than a factor of five larger.
}
Due to the presence of both La and Br $d$-states as well La-$f$ states, the density of states in the conduction band is larger than in conventional semiconductors such as Si or GaAs. As a result, one needs five times as many unoccupied than occupied states in order to reach the convergence exhibited in \fig{fig:qp_conv}(a), a number that is considerably larger than for the aforementioned semiconductors. In general we have found it very useful not only to check convergence with respect to the number of bands as shown in \fig{fig:qp_conv}(b) but also to confirm convergence of our calculations by comparing the results obtained from initially very different wave functions such as DFT and EXX.

\subsection{Hybrid DFT and $G_0W_0$ calculations}
\label{sect:results_hybrid}

\begin{figure}
  \centering
\includegraphics[scale=0.62]{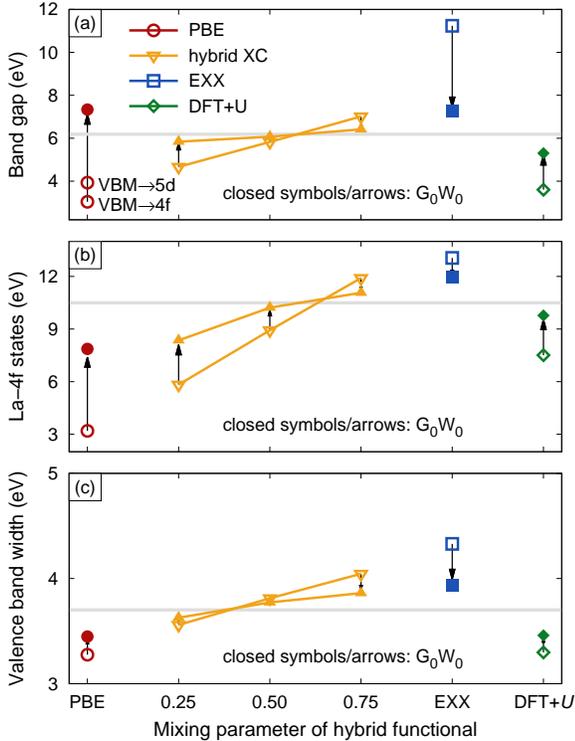}
  \caption{
    (a) Band gap, (b) position of La-$4f$ states, and (c) valence band width from conventional exchange-correlation functional (PBE), several hybrid exchange-correlation functional (hybrid XC), and exact-exchange (EXX) calculations. Vertical arrows indicate the improvement obtained by carrying out $G_0W_0$ calculations. Gray bars indicate QPscGW values that serve as reference data in this comparison. All data obtained without spin-orbit coupling.
In (a) two values are shown for PBE corresponding to the gaps between VBM and La-$4f$ and $5d$ states, respectively (compare \fig{fig:dostot}). For all other cases the band gap between VBM and La-$5d$ is shown.
  }
  \label{fig:levels}
\end{figure}

For studying various problems of interest such as charge self-trapping, defect formation, or alignment of activator levels one needs to consider ionic relaxations and representative supercells. At present QPscGW calculations are, however, computationally still extremely expensive because a large number of unoccupied bands has to be included. More severely, the method does not allow to obtain total energies and forces. It is therefore important to determine to which extent computationally lesser demanding calculation schemes can reproduce the reference electronic structure provided by QPscGW calculations. To this end, a number of conventional and hybrid exchange-correlation functionals, as well as the corrections from $G_0W_0$ to some of these functionals have been considered.

Figure~\ref{fig:levels} shows the variation of several QP energy differences with the mixing parameter $\alpha$ of a range-separated hybrid exchange-correlation functional\cite{HeyScuErn03} with a screening length of $\mu=0.2$\, \AA$^{-1}$. For $\alpha=0$ this functional is identical to the XC functional by Perdew \etal\ (PBE). \cite{PerBurErn96} For comparison we also included data from EXX calculations, which do not include correlation.

The data shows that at the PBE level, the ordering of the La-$4f$ and $5d$ states is incorrect with $4f$ located below $5d$ levels. Upon inclusion of exact-exchange and/or by adding a $G_0W_0$ calculation the level ordering is corrected but for most functionals the QP energies are still quantitatively very different from the QPscGW values. In general as $\alpha$ is increased the QP energy differences considered here increase. Performing $G_0W_0$ calculations in general improves the agreement. The same trend was observed by Aulbur {\em et al.} for medium and wide-gap materials using $G_0W_0$ on top of an EXX-based XC functional.\cite{AulStaGor00} While the $G_0W_0$ method reduces the dependence of the results on $\alpha$, it remains pronounced.

\begin{figure}
  \centering
\includegraphics[scale=0.62]{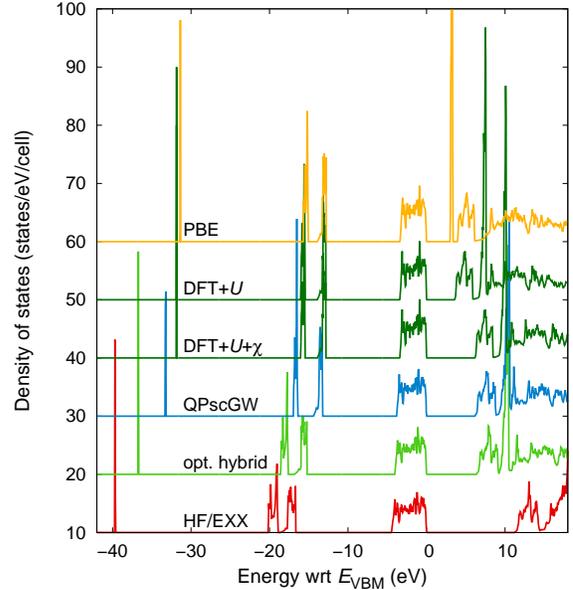}
  \caption{
    Density of states as obtained from various different calculation schemes (without SOC).
  }
  \label{fig:dostot}
\end{figure}

For $\alpha$ values between approximately 0.5 and 0.65 all QP energy differences considered in \fig{fig:levels} intersect the respective reference values. This suggests that a hybrid functional with $\alpha$ in this range could possibly reproduce the DOS obtained from QPscGW calculations. We therefore considered a hybrid functional with $\alpha=0.62$, for which the total DOS is shown in \fig{fig:dostot} together with results obtained from PBE, EXX, DFT+$U$, and QPscGW. For simplicity SOC effects were neglected in this comparison. In fact apart from the deep semi-core states the DOS for the modified hybrid functional compares very favorably with the QPscGW reference.

The figure also contains the DOS obtained from DFT+$U$ calculations using $U$ and $J$ parameters determined previously for other La-compounds, \cite{CzySaw94} both as-calculated (DFT+$U$) and with a rigid shift of the conduction bands to the QPscGW band gap (``scissors'' correction, DFT+$U+\chi$). Both the upper valence and lower conduction band structures from DFT+$U+\chi$ are in excellent agreement with QPscGW data, and even though the agreement worsens for deeper lying valence states, among the methods considered the DFT+$U+\chi$ approach still yields the best agreement with the reference DOS. Still for practical calculations one would have to resort to the DFT+$U$ method without the scissors correction, which exhibits a pronounced band gap underestimation (3.60\,eV vs 6.19\,eV without SOC).

\begin{figure}
  \centering
\includegraphics[scale=0.62]{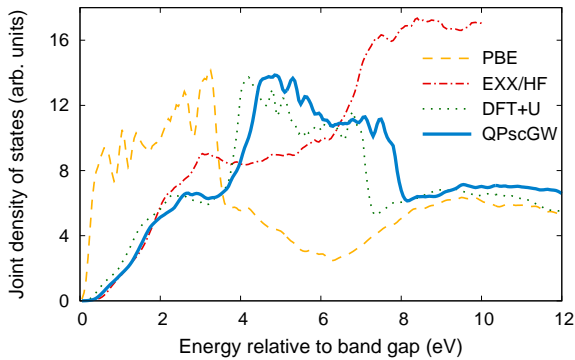}
  \caption{
    Joint density of states from different calculation schemes (without SOC). It is instructive to compare these curves with the data in \fig{fig:jdos_abs}.
  }
  \label{fig:jdos_comp}
\end{figure}

In this context it is furthermore interesting to revisit the JDOS, shown in \fig{fig:jdos_comp}, which was compared earlier with absorption data (see \fig{fig:jdos_abs}). Also in this arena both the DFT+$U$ method and the modified hybrid XC functional (not shown in \fig{fig:jdos_comp}) yield good agreement with QPscGW calculations exhibiting the features discussed in detail in \sect{sect:results_qpscgw}. In stark contrast the JDOS from PBE and EXX calculations are qualitatively very different from the QPscGW data. The position of the features in the absorption spectra shown in \fig{fig:jdos_abs} are not reproduced with these functionals.

\section{Discussion and conclusions} 
\label{sect:conclusions}

In this paper it has been demonstrated that QPscGW calculations of LaBr$_3$ yield single particle spectra (DOS) that closely match experimental XPS data (\fig{fig:band_qpscgw}). At the same time the key features in the JDOS correlate with those in the experimental absorption spectrum (\fig{fig:jdos_abs}). To obtain this result it was essential to include spin-orbit coupling in the calculations. The very good agreement between our QPscGW calculations and experimental data provides confidence in the reliability of the QPscGW method for treating systems with $f$-electrons and weak to moderate correlation. Our work complements earlier studies of other $f$-electron systems that used both the QPscGW method \cite{SchKotFal06,ChaSchKot07} and the $G_0W_0$ approach \cite{JiaGomRin09} but focused on systems, in which $f$-electrons are located in the valence band.

The presence of both $f$ and $d$ states in the conduction band of LaBr$_3$ requires ---by comparison with more conventional semiconductors and insulators--- an unusually large basis set for the expansion of the effective Hamiltonian to obtain converged results [\fig{fig:qp_conv}(b)]. It was also explicitly demonstrated that the QPscGW results are independent from the initial wave function [\fig{fig:qp_conv}(a)]. This property is a big advantage of the QPscGW approach compared to other $GW$ methods, as exemplified in \fig{fig:levels}, which shows the $G_0W_0$ results to exhibit a rather pronounced dependence on the initial wave function. This behavior is important to keep in mind when interpreting such calculations.

Another aspect that deserves mentioning is related to the computation of matrix elements and their interpolation. The QPscGW method does provide updated wave functions that in turn can be used to compute matrix elements. In contrast the wave functions are unchanged when doing $G_0W_0$ calculations. This difference becomes particularly apparent when interpolation methods are used. These methods, based for example in the present work on maximally localized Wannier functions, can be used to represent the aforementioned matrix elements on a finer mesh in reciprocal space, dramatically improving convergence of the calculations. If, as in the case of LaBr$_3$, the level ordering changes as the result of a $G_0W_0$ calculation the connectivity of the states changes as well, which causes problems for interpolation methods. In such a case the interpolation of the QP energies and the projections on the basis of a $PBE+G_0W_0$ calculation are physically not meaningful because the derivative matrices $M^{(\vec{k},\vec{b})}$ no longer refer to the same states. The QPscGW method, however, does not suffer from this shortcoming and matrix elements can be readily interpolated once the MLWFs have been determined.

It is instructive to compare the electronic structure of LaBr$_3$ both with other lanthanum halides and the free La atom. As discussed at length in this paper in LaBr$_3$ the La-$4f$ levels are located {\em above} the La-$5d$ levels. This is similar to the ordering of {\em excited} states in a free La$^{2+}$ ion. \cite{MarZalHag78} In contrast for the $3+$ charge state of the La ion the excited $4f$ states are indeed observed to lie below the $5d$ levels. This situation is compatible with a partial charge transfer between La and Br.

Finally, it was shown that a hybrid XC functional can be constructed that yields reasonable agreement with DOS (\fig{fig:dostot}) and JDOS (\fig{fig:jdos_comp}) obtained from QPscGW calculations. This functional should be suitable for studies that require ionic relaxation and/or the use of supercells. The DFT+$U$ method (using parameters from the literature \cite{CzySaw94}) also reproduces the DOS structure of valence and conduction band states independently but still underestimates the band gap significantly.

\begin{acknowledgments}
This work was performed under the auspices of the U.S. Department of Energy by Lawrence Livermore National Laboratory under Contract DE-AC52-07NA27344 with support from the National Nuclear Security Administration Office of Nonproliferation Research and Development (NA-22). One of us (PE) acknowledges partial support through the ``Area of Advance: Materials'' at Chalmers University of Technology.
\end{acknowledgments}

\end{document}